\documentclass[12pt]{article} 
\usepackage{epsfig}
\usepackage{amsfonts}
\usepackage{latexsym}
\usepackage{amsmath,amssymb}
\usepackage{verbatim}
\usepackage{setspace}
\usepackage{tikz}
\usepackage[dvipdfm,hypertex]{hyperref}
\usepackage[textheight=9in, textwidth=6.5in, letterpaper]{geometry}

\numberwithin{equation}{section}

 \def\p{\partial}

\newcommand{\bea}{\begin{eqnarray}}
\newcommand{\eea}{\end{eqnarray}}
\newcommand{\be}{\begin{equation}}
\newcommand{\ee}{\end{equation}}
\newcommand{\ba}{\begin{align}}
\newcommand{\ea}{\end{align}}

\newcommand{\bigO}{\mathcal{O}}

\newcommand{\is}{i_1 \cdots i_s}
\newcommand{\scriplus}{\mathcal{I}^+}
\renewcommand{\epsilon}{\varepsilon}

\newcommand{\chichi}{\chi\!\cdot\!\chi}

  \makeatletter
  \let\over=\@@over \let\overwithdelims=\@@overwithdelims
  \let\atop=\@@atop \let\atopwithdelims=\@@atopwithdelims
  \let\above=\@@above \let\abovewithdelims=\@@abovewithdelims

\renewcommand\section{\@startsection {section}{1}{\z@}%
                                   {-3.5ex \@plus -1ex \@minus -.2ex}
                                   {2.3ex \@plus.2ex}%
                                   {\normalfont\large\bfseries}}

\renewcommand\subsection{\@startsection{subsection}{2}{\z@}%
                                     {-3.25ex\@plus -1ex \@minus -.2ex}%
                                     {1.5ex \@plus .2ex}%
                                     {\normalfont\bfseries}}

\def\a{\alpha}
\def\b{\beta}

\def\Or[#1]{{\text{O}}\left({#1}\right)}
\def\dotl[#1,#2]{\left\langle #1, #2 \right\rangle}
\def\dotlb[#1,#2]{[ #1, #2 ]}
\def\dotp[#1,#2]{(#1) \cdot (#2)}
\def\aff[#1,#2]{\hat{#1}(#2)}
\def\n4sym{{\cal N}=4 SYM}
\def\>{\rangle}
\def\<{\langle}
\def\weight[#1,#2,#3]{\{(#1),#2,#3\}}
\def\ads[#1]{$\text{AdS}_{#1}$}

\def\vx{{\vec x}}

\def\p{\partial}

\begin{document}
\unitlength = 1mm
\ \\

\begin{center}

{ \LARGE {\textsc{ Higher Spin  Realization of the \\ \vspace{.5cm} DS/CFT Correspondence}}}

\vspace{0.8cm}

Dionysios Anninos$^a$, Thomas Hartman$^b$ and Andrew Strominger$^c$

\vspace{1cm}

\vspace{0.5cm}

{\it $^a$ Stanford Institute for Theoretical Physics, Stanford University,\\
Stanford, CA 94305, USA}

\vspace{0.5cm}

{\it $^b$ Institute for Advanced Study, School of Natural Sciences,\\
Princeton, NJ 08540, USA}

\vspace{0.5cm}

{\it $^c$ Center for the Fundamental Laws of Nature, Harvard University,\\
Cambridge, MA 02138, USA}

\vspace{0.5cm}

\vspace{1.0cm}

\end{center}

\begin{abstract}
\noindent We conjecture that Vasiliev's theory of higher spin gravity in four-dimensional de Sitter  space (dS$_4$) is holographically dual to a three-dimensional conformal field theory (CFT$_3$) living on the spacelike boundary of dS$_4$ at future timelike infinity.  The CFT$_3$ is the Euclidean $Sp(N)$ vector model with anticommuting scalars.  The free CFT$_3$ flows under a double-trace deformation to an interacting CFT$_3$ in the IR. We argue that both CFTs are  dual to  Vasiliev dS$_4$ gravity but with different future boundary conditions on the bulk scalar field.  Our analysis rests heavily on analytic continuations of bulk and boundary correlators in the proposed duality relating the $O(N)$ model with Vasiliev gravity in AdS$_4$.
\end{abstract}

\pagestyle{empty}

\pagebreak 
\setcounter{page}{1}
\pagestyle{plain}

\setcounter{tocdepth}{1}

\tableofcontents

\section{Introduction}

The AdS/CFT correspondence provides a non-perturbative holographic definition of anti-de Sitter (AdS) quantum gravity in terms of a CFT living on the timelike conformal boundary of AdS.  Our own universe is unlikely  to  have an anti-de Sitter  boundary, but may well have a de Sitter (dS) boundary in the far future.  This dS boundary shares a number of mathematical properties with the AdS boundary. Hence it is natural to try to define dS quantum gravity in terms of a CFT living on the future conformal boundary of  dS \cite{Strominger:2001pn,Hull:1998vg,Witten:2001kn,Strominger:2001gp,Balasubramanian:2002zh, Maldacena:2002vr}.  
One  key difference is that in AdS/CFT, the radial direction emerges holographically from the CFT, while in dS/CFT time itself must be holographically emergent. It is challenging  to reconcile this with our usual quantum notions of unitary time evolution. A second key difference is that we have had no
useful microscopically complete  examples of the dS/CFT correspondence. This has stymied progress in the subject and at times rendered the discussions somewhat formal.\footnote{Although it is important to note that universal properties of the dS/CFT correspondence, as applied to the dS era proposed by inflation, nevertheless appear useful for analysis of CMB data
\cite{Maldacena:2002vr}.} 
It is the purpose of this paper to begin to fill this gap.

Specifically,  we conjecture that Vasiliev's higher spin gravity \cite{Vasiliev:1990en,Vasiliev:1999ba}  in dS$_4$ is holographically dual to the  three-dimensional conformal field theory (CFT$_3$) with anticommuting scalars and $Sp(N)$ symmetry studied by LeClair and collaborators in \cite{LeClair:2006kb,LeClair:2006mb,LeClair:2007iy,Robinson:2009xm}.  This is a de Sitter analogue of the conjectured Giombi-Klebanov-Polyakov-Yin  (GKPY) duality relating the $O(N)$ CFT$_3$ to Vasiliev gravity in AdS$_4$, whose remarkable properties have received much recent attention \cite{Klebanov:2002ja, Witten,Mikhailov:2002bp,Sezgin:2002rt, Giombi:2009wh,Giombi:2010vg,Giombi:2011ya}. The $Sp(N)$ CFT$_3$ dual to de Sitter space has anticommuting scalar fields and is therefore non-unitary.  This peculiarity does not rule out the duality because in dS/CFT, the CFT is Euclidean and never continued to Lorentzian signature. On the other hand, the good properties of ordinary time evolution in the bulk must be encoded somehow in the CFT.  Indeed the $Sp(N)$ CFT$_3$  turns out to have a  ``pseudo-unitary"  structure \cite{LeClair:2007iy} which may be relevant.

We consider the parity-invariant minimal bosonic version of Vasiliev  gravity, which has massless symmetric tensor fields of all even spins $s=0,2,4,\dots$.  Versions of this theory exist in both Euclidean 
AdS$_4$ (EAdS$_4$)  and dS$_4$, both of which have an $O(4,1)$ symmetry. 
These  EAdS$_4$  and dS$_4$ versions turn out to be very simply related to one another by an analytic continuation which reverses the sign of the cosmological constant $\Lambda$ while Newton's constant $G_N$ is held fixed.\footnote{This is a very special feature of Vasiliev gravity. No such nice continuations are known for AdS solutions of string theory: one typically encounters imaginary conformal weights, imaginary fluxes, and negative energy or ghost fields in the bulk. }  In the GKPY duality, $N\sim  \frac{1}{\Lambda G_N} $, so this operation takes $N\to -N$. 
The appearance of $Sp(N)$ can then be understood as an analytic continuation of $O(N)$ under $N \to -N$. We will show below that singlet correlators of the $Sp(N)$ model are obtained by replacing $N$ with $-N$ in the expressions for the corresponding $O(N)$ correlators.   This relation holds because the closed loops which give powers of $N$ are always accompanied by an extra minus sign from Fermi statistics in the $Sp(N)$ theory. 

GKPY duality comes in two flavors. In the first, Vasiliev AdS$_4$ gravity with Neumann boundary conditions on the scalar field is related to the free $O(N)$ CFT$_3$.  In the second, Vasiliev AdS$_4$ gravity with Dirichlet  boundary conditions on the scalar field is related to the critical interacting $O(N)$ CFT$_3$. These two GKPY dualities lead to two corresponding dS/CFT dualities, which relate Vasiliev dS$_4$ gravity with Neumann (Dirichlet) scalar boundary conditions at future infinity to the free (critical interacting) $Sp(N)$ CFT$_3$. All evidence for either GKPY duality analytically  continues to evidence for the corresponding dS/CFT duality. 

While the $n$-point corrrelators are the basic building blocks of the dS/CFT correspondence, there is much to be said and understood about how they are related to physical observables. In \cite{Maldacena:2002vr} it was shown that the generating function of these correlators can be viewed as  a Wheeler-DeWitt or Hartle-Hawking type wave function of the universe.   There have been a number of interesting recent developments along theses lines \cite{Anninos:2009yc,Anninos:2010gh,Anninos:2010zf, Anninos:2011vd,Harlow:2011ke,
Castro:2011xb, Maldacena:2011nz,McFadden:2011kk, Maldacena:2011mk,  Anninos:2011jp} which clarify the structure of dS/CFT. At the same time, important conceptual issues remain including the interpretation of the dS entropy and the nature of physical observables and bulk unitarity. The correlators we compute reside at $\mathcal{I}^+$ and are thus `metaobservables', inaccessible to any single observer. On the other hand, the de Sitter entropy is naturally associated to a static patch observer in de Sitter space (see \cite{Gibbons:1977mu,Goheer:2002vf,Banks:2006rx,Dong:2010pm,Silverstein:2003jp}), whose worldline only intersects $\mathcal{I}^+$ at a single point. In this paper, we will steer clear of all these important issues and focus on the mathematical relationship between the bulk gravity and boundary CFT correlators. 
 Nevertheless, we hope that the concrete example of dS/CFT we present below will prove useful for future discussions.

\section{CFT$_3$ correlators}

\subsection{Free theory}
The free $Sp(N)$ model is described by the action 
\be\label{freeaction}
I_{Sp(N)}^{free} = \frac{1}{8\pi} \int d^3 x \ \Omega_{ab} \delta^{ij} \p_i \chi^a \p_j \chi^b  \ , \quad \quad \Omega_{ab} = \left(\!\begin{array}{cc}
0 & 1_{\frac{N}{2}\times\frac{N}{2}}\\
-1_{\frac{N}{2}\times\frac{N}{2}} & 0
\end{array}\!\right)
\ee
where $\chi^a$ is a real $N$-component anticommuting scalar, $\Omega_{ab}$ is the antisymmetric symplectic form, and $N$ is even. The action is invariant under $Sp(N)$ rotations of $\chi$. This theory has been studied in a variety of contexts: a similar $N=2, d=2$ logarithmic CFT with central charge $c=-2$ describes polymers \cite{Saleur:1991hk,Flohr:1995ea,Kausch:1995py}, and the $d=3$ model of interest for dS/CFT was considered in the context of superconductivity and critical phenomena in \cite{LeClair:2006kb,LeClair:2006mb,LeClair:2007iy}.  

The free theory  has singlet conserved currents of all even spins:
\be
J^{(s)}_{\is} = \Omega_{ab}\chi^a \p_{(i_1}\cdots \p_{i_s)} \chi^b + \cdots \ .
\ee
Here normal ordering is implied, $a,b=1,\cdots N$  and the corrections render the currents traceless and conserved.
$J^{(s)}$ has dimension $\Delta = s+1$, and in particular the scalar operator,
\be
J^{(0)} = \chichi \equiv \Omega_{ab}\chi^a\chi^b\ ,
\ee
has $\Delta=1$.  We would like to show that free $Sp(N)$ correlators are related to free $O(N)$ correlators\footnote{ In our notation for $Sp(N)$, $N$ is always even, e.g. $SU(2) \approx Sp(2)$. Various properties of the groups (for example character formulae) are related under analytic continuation by \cite{Cvitanovic:1982bq,Dunne:1988ih,Mkrtchyan:2010tt} 
$O(-N) = \overline{Sp(N)}$
where the bar indicates a transpose of all Young tableaux. The transpose swaps symmetric and antisymmetric representations, which is accomplished in the CFT by swapping bosons for fermions.} by $N\rightarrow -N$.

 This is easy because the free connected $n$-point functions are just one loop diagrams with $n$ current insertions around the loop. The only $N$ dependence (for explicit expressions see \cite{Giombi:2009wh}) is in the overall factor, so we need connected correlators to obey
\be\label{signflip}
\langle J^{(s_1)}\cdots J^{(s_n)}\rangle_{Sp(N)}^{free} = - \langle J^{(s_1)}\cdots J^{(s_n)}\rangle_{O(N)}^{free} \ .
\ee
 The minus sign is the usual $(-1)$ for each loop from Fermi statistics. 
To see this factor explicitly, we write  $\chi = (\psi_1, \psi_2)$ where $\psi_{1,2}$ are real ${N}/{2}$-component anticommuting scalars. The currents are
\be
J^{(s)}_{\is} = \delta_{\a\b}\psi_1^\a \p_{(i_1}\cdots \p_{i_s)} \psi_2^\b  +  \cdots - (\psi_1 \leftrightarrow \psi_2) \ 
\ee
where $\a,\b=1,\cdots {N \over 2}$. Wick contractions of $\psi_{1,2}$ then lead immediately to (\ref{signflip}).

\subsection{Interacting theory}
We now consider the interacting $Sp(N)$ model
\be\label{cftlag}
I_{Sp(N)} = \frac{1}{8\pi} \int d^3 x \left( \Omega_{ab} \delta^{ij} \p_i \chi^a \p_j \chi^b + m^2 \chichi + \lambda(\chichi)^2 \right) \ .
\ee
Like the interacting $O(N)$ model, this theory has an interacting IR fixed point\cite{LeClair:2007iy}. This defines the critical model, reached by flowing to $\lambda_{IR}$ and tuning the renormalized mass to zero.
It has been studied in  \cite{LeClair:2007iy} in the $\epsilon$ expansion to two loops, where it was found that $\beta$ functions and anomalous dimensions are related to the critical $O(N)$ results by $N\rightarrow -N$.  Our interest here is restricted to correlation functions of the $Sp(N)$ singlet operators $J^{(s)}$.  For such operators one may show diagrammatically that, as for the free theory, the correlators are related to their $O(N)$ counterparts simply by the replacement of $N\to -N$.  Relative minus signs arise in the $Sp(N)$ theory because the scalars anticommute, but these are correlated with the powers of $N$. 

The simplest way to see this in detail is to introduce a (commuting) Lagrange multiplier field $\alpha$ and rewrite (\ref{cftlag}) in the form \cite{LeClair:2007iy}
\be\label{cftag}
I_{Sp(N)} = \frac{1}{8\pi} \int d^3 x \left( \Omega_{ab} \delta^{ij} \p_i \chi^a \p_j \chi^b + m^2 \chichi + 2\alpha \sqrt{\lambda}(\chichi)-\alpha^2 \right) \ .
\ee
The Feynman diagrams for singlet correlators in this theory consist solely of closed loops of 
anticommuting $\chi$ scalars connected by $\alpha$ propagators. Powers of $N$ come only from closed $\chi$ loops, and each of these loops gives one minus sign from Fermi statistics. 
Hence these diagrams are in one-to-one correspondence with those of the $O(N)$ model and 
differ only in the replacement of $N$ by $-N$.  
This relation holds for all values of $\lambda$, including the critical IR fixed point of interest to us.  
To confirm that this perturbative argument applies even at the fixed point, it is straightforward to compute anomalous dimensions in the $1/N$ expansion by modifying the $O(N)$ computations in \cite{Wilson:1972cf}. The result agrees with the $O(N)$ model after $N \rightarrow -N$, at least through order $1/N$.\footnote{Curiously, upon developing an $\epsilon$-expansion, it was argued in \cite{LeClair:2006kb} that the critical value of the coupling constant $\lambda$ of the $Sp(N)$ theory becomes negative for $N>4$.}

\subsection{Summary}

  The singlet correlators of the  free (critical interacting) $Sp(N)$ model, are obtained by the replacement of $N$ with $-N$ in the corresponding singlet correlators of the  free (critical interacting) $O(N)$ model.

\section{Bulk correlators}\label{ss:ancon}

On the bulk side we consider Vasiliev's minimal bosonic higher spin theory  in dS$_4$ \cite{Vasiliev:1990en,Vasiliev:1999ba}. The representation of the dS$_4$  solution in terms of Vasiliev's master field, and its relation by analytic continuation to EAdS$_4$,  is described  in \cite{Iazeolla:2007wt} to which we refer the reader for details. 
The dS$_4$ metric is the only nonzero background and takes the form, in planar coordinates
\be\label{dsf}
ds^2 =\frac{ \ell_{dS}^2}{\eta^2}\left(-d\eta^2 + (d\vx)^2\right) \ ,
\ee
where $\eta\le 0$ and $\vx$ are coordinates on the planar $R^3$ slices. In these coordinates 
the future boundary $\scriplus$ is at $\eta=0$  while the past horizon is at $\eta=-\infty$. 
The spectrum of conformal dimensions for linearized excitations about this solution agrees with the $Sp(N)$ model:  there are massless symmetric tensor fields of all even spins $s$, with scaling dimensions $\Delta = s+1$.  Note that all scaling dimensions are real, which is not in general guaranteed in dS/CFT.

We now compute the boundary correlators of the Vasiliev theory in dS$_4$, in order to compare to the $Sp(N)$ CFT. This could be done directly in dS$_4$, but it is simpler to borrow results from the Vasiliev theory in AdS$_4$, where 3-point correlation functions have already been computed by Giombi and Yin \cite{Giombi:2009wh,Giombi:2010vg,Giombi:2011ya}.  We will show that Vasiliev $n$-point correlators in dS$_4$ can be obtained from those in EAdS$_4$ by analytically continuing $N \to -N$.

The full nonlinear equations of motion are formulated in terms of master fields that are functions of the spacetime coordinates $(\vx,\eta)$ and valued in the higher spin algebra, which is a quotient of $U(SO(5,C))$.  The equations of motion for the master fields take the same form for positive or negative $\Lambda$ in either signature, but this choice enters as a reality condition on the master fields, or equivalently a choice of real form for the higher spin algebra:\footnote{This is similar to the Chern-Simons formulation of 3d gravity \cite{Achucarro:1987vz,Witten:1988hc}, where both Lorentzian dS$_3$ gravity and Euclidean AdS$_3$ gravity are described by $G_C = SL(2,C)$ gauge theory.  In the former case, $G_C$ is considered to be the complexification of $G=SL(2,R)$, while in the latter case it is treated as the complexificaton of $G=SU(2)$ \cite{Witten:1989ip}. } see  \cite{Iazeolla:2007wt}. The equations of motion for the infinite tower of real fields $X= \varphi, h_{\mu\nu}, w_{\mu\nu\rho\sigma}, \dots$ of spins $0,2,4,\dots$ can be extracted order by order from the Vasiliev equations, and will depend on $\Lambda$ and the signature. The linearized equation of motion for a spin-$s$ field is, schematically,
\be\label{bulkfree1}
\nabla^2 X + \Lambda  X=0~.
\ee
In general, the corrections will be an expansion in $\frac{1}{\Lambda}\nabla^2$. For example, cubic interactions among spins $(2,s,s)$ involving the spin 2 metric fluctuation $h$ given explicitly in \cite{Kristiansson:2003xx,Boulanger:2008tg} take  the schematic form
\be\label{cantwoeom}\nabla^2h\sim \Lambda X^2 +X\nabla^2 X
+(\nabla X)^2 + \frac{1}{\Lambda}( \nabla^3 X \nabla X + (\nabla^2 X)^2) +\cdots \ .
\ee
Although we have dropped the constants and indices in these schematic expressions, all factors of $\Lambda$ are written.  Note that we have taken the $bulk$ dimension of $X \sim ({\rm length})^0$ to vanish, while $\nabla\sim  ({\rm length})^{-1}, ~\Lambda\sim ({\rm length})^{-2}$.

The  Poincar\'{e} patch metric of  EAdS$_4$ is 
\be
ds^2 =\frac{ \ell_{AdS}^2}{z^2}\left(dz^2 + dx^i dx^i\right) \ ,
\ee
where $z$ is the radial coordinate.
This is obtained from the planar patch dS$_4$ metric (\ref{dsf}) by the double Wick rotation
\be\label{ancon}
 \quad z = - i \eta \ , \quad \ell_{EAdS} = -i \ell_{dS} \ .
\ee
This rotation transforms the equations of motion (\ref{bulkfree1}) and (\ref{cantwoeom}) from dS$_4$ to
EAdS$_4$.  It is perhaps simplest to work in a generalized Fefferman-Graham (FG) gauge near the boundary, in which all components of the higher spin fields with indices tangent to the radial direction (e.g. $w_{ijkz}$) are set to zero. In this case all components of $X$ are tangent to the boundary. The relation between the reality condition on the master fields  in EAdS$_4$ and dS$_4$ imply that the FG gauge components of $X$ are kept real under the continuation. This is shown in detail in appendix \ref{app:realx}. 

Now we would like to compare the $n$-point correlators in EAdS$_4$ to those of dS$_4$.  These are not fully determined by the classical equations of motion. One extra dimensionless parameter is required which governs the strength of quantum corrections. In a theory with an action, this parameter is essentially the overall coefficient of the action, roughly  $N\sim {1 \over G_N\Lambda }$. As shown in \cite{Giombi:2009wh}, in the language of correlators adopted here, the extra parameter can be identified with the dimensionless coefficient $C_2$ of the tree-level two point function of the bulk scalar $\varphi$.   A rule must be given for the behavior of $C_2$ under the analytic continuation (\ref{ancon}). If we had an action, this rule would be determined from positivity of the energy in Lorentzian signature. In the language of correlators, the rule is equivalently fixed by demanding that the short distance singularity has the canonical Hadamard sign. 
At short distances $(z_1,\vx_1)\to (z_2, \vx_2)$ in EAdS$_4$ 
\be\label{sds} \langle \varphi(\vx_1,z_1)\varphi(\vx_2,z_2)\rangle_{EAdS, tree} \sim {C_2 z_1z_2\over (z_1-z_2)^2+ x_{12}^2} \ , \quad x_{12}^2 \equiv (\vec{x}_1 - \vec{x}_2)^2 \ , \ee 
where $C_2$ is positive. If we take $z\to-i\eta$ in (\ref{sds}) without changing the sign of $C_2$, we get the wrong sign in Lorentzian signature. Therefore in continuing to dS$_4$ we must supplement  (\ref{ancon}) with
\be\label{dsz}  C_2\to -C_2.\ee
Note that since roughly speaking  $C_2\sim -G_N\Lambda \sim {1 \over N} $ in EAdS$_4$, we  can think of this as taking $\Lambda\to -\Lambda$ with $G_N$ fixed or equivalently $N\to -N$.

Finally putting this together gives a rule for the continuation of bulk correlators. 
Holding the field $X$ fixed while performing the rotation (\ref{ancon})(\ref{dsz}), the bulk $n$-point functions transform as 
\be\label{rle}
   \langle X(\eta_1,\vx_1)\cdots X(\eta_n,\vx_n) \rangle_{dS}=  \langle X(iz_1,\vx_1)
   \cdots X(iz_n,\vx_n) \rangle_{EAdS}|_{\Lambda \to -\Lambda}^{C_2\to -C_2}.
   \ee
 In summary, to get the bulk dS$_4$ correlators,  we simply take the EAdS$_4$ correlators and  replace 
   $z$ with $-i\eta$, $\Lambda$ with $-\Lambda$ and $C_2$ with $-C_2$.  In Appendix A, we rederive this rule in the on-shell action formalism.

\subsection{Spin zero corrrelators}
To compare with the boundary CFT$_3$ correlators, we must take the arguments of the bulk correlators to the boundary $z=z_c\to 0$ and rescale.  Let us first consider the bulk  scalar two point function.   In EAdS$_4$, a scalar of dimension
$h_+=\frac{3}{2}+\sqrt{\frac{9}{4}+m^2\ell_{AdS}^2}$ behaves as $\varphi(\vx,z) \sim \varphi_+(\vx) z^{h_+} + \varphi_-(\vx) z^{3-h_+}  + \cdots$ near the boundary $z\to 0$.  The bulk scalar $\varphi$ has $h_+=2$ but it is instructive to consider the more general case. 
To compute the correlators a boundary condition must be specified. To be specific let us consider the case $\varphi_-=0$. 
Dimensional analysis and $O(4,1)$ invariance then imply that at  tree-level\footnote{ The exact tree-level two point function is (see for example \cite{D'Hoker:1999ni}) $C_2({1\over2(P-1)})^{h_+}F(h_+,h_+-1,2h_+-2;-{2 \over P-1})$ with $P=\left(z_1^2+z_2^2+x_{12}^2\right)  /2z_1z_2$, so that with the normalization used here a positive
constant $2^{2h_+}\Gamma(h_+)\Gamma(h_+-1)/(4\pi)^2\Gamma(2h_+-2)$ appears on the right hand side of (\ref{sds}).}\be \label{eadsbdy} \langle \varphi(\vx_1,z_c)\varphi(\vx_2,z_c) \rangle_{EAdS, tree}=C_2\left({z_c^2\over x_{12}^2}\right)^{h_+} \ .\ee
Dimensional analysis and $O(4,1)$ invariance similarly imply for the full  connected $n$-point functions 
\be \label{npt} \langle \varphi(\vx_1,z_c)\varphi(\vx_2,z_c)...\varphi(\vx_n,z_c) \rangle_{EAdS}=A_n (C_2)^{n} \left({z_c\over x}\right)^{nh_+},\ee
where by $x$ we just mean something which scales linearly under rescalings of all $n$ $\vx_i$s. Since these are unaffected by the analytic continuation we do not need to keep track of the detailed form. Note that in general the constants $A_n=A_n[C_2]$. The $g$-loop contribution to $A_n$  goes as $ (C_2)^{g-1}$ i.e.
\be A_n[C_2]=\sum_{g=0}^\infty A_n^{(g)} (C_2)^{g-1}\ee  where $A_n^{(g)}$ is in principle determined from the classical equation of motion and does not depend on $C_2$. 
Defining 
\be \mathcal{O} =\left({1 \over C_2 z_c^{h_+}}\right)\varphi \ee
we get the boundary correlators
\be \langle \mathcal{O}(\vx_1)...\mathcal{O}(\vx_n)\rangle_{EAdS}= {A_n [C_2] }x^{-nh_+}.\ee
In the bulk classical limit of small $C_2$, these connectd $n$-point functions all scale as $C_2^{-1}$.

This is the answer in the EAdS/CFT correspondence.  To transform to dS$_4$, we go back to the bulk correlators and use the analytic continuation (\ref{rle}). Noting that $h_+$ is preserved under  (\ref{ancon}), we obtain \be \label{npt} \langle \varphi(\vx_1,\eta_c)\varphi(\vx_2,\eta_c)...\varphi(\vx_n,z_c)\rangle_{dS}=A_n[-C_2] (- C_2)^n \left({-i\eta_c\over x}\right)^{nh_+},\ee
Defining 
\be \mathcal{O}=\left({1 \over - C_2 (-i\eta_c)^{h_+}}\right)\varphi \ee
we get the boundary correlators
\be\label{dsl} \langle \mathcal{O}(\vx_1)...\mathcal{O}(\vx_n)\rangle_{dS}= A_n[-C_2] x^{-nh_+}.\ee
These correlators obey future boundary conditions which are the analytic continuation of 
the EAdS$_4$ ones.\footnote{The appearance of correlators defined by future boundary conditions may at first sound odd. This may be motivated from the dS perspective either as an intermediate device for computing 
Wheeler-DeWitt type wave/generating functions \cite{Maldacena:2002vr}, or as in \cite{Anninos:2011jp} from the fact that they do not lead to observable causality violation: operationally there is no difference. See appendix \ref{app:action} for further discussion.}
Note however that the boundary correlators are  defined intrinsically in dS$_4$; we are just using EAdS$_4$ as a computational tool.

\subsection{Higher spin correlators}
This result readily extends to the correlators of the higher spin fields. The detailed structure in EAdS$_4$ is 
spelled out in \cite{Giombi:2009wh} and will not be reviewed here. In FG gauge the  higher spin fields are then in one-to-one correspondence with the  currents of the boundary CFT$_3$.\footnote{Reference \cite{Giombi:2009wh} employs a divergence-free gauge in which $X$ has more components, but there are relations among them. The analytic continuation procedure is slightly more involved in divegence-free gauge but the final results are equivalent.}
Retracing the steps of the previous section leading to (\ref{dsl}), we again conclude that the dS$_4$ correlators are obtained from the EAdS$_4$ correlators simply by replacing $C_2$ with $-C_2$.

\section{Bulk-boundary  comparison}

   In section 2 we showed that the $n$-point functions of the free  (critical) 
   $Sp(N)$ model are those of the free  (critical)  $O(N)$ model with $N$ replaced by $-N$.
   In section 3 we showed that the rescaled boundary correlators of Vasiliev
   dS$_4$ gravity with the Neumann (Dirichlet) scalar boundary condition  are those of Vasiliev  EAdS$_4$ gravity with the Neumann (Dirichlet)  scalar boundary condition with $C_2$ replaced by $-C_2$.  According to the conjectured GKPY duality, the corrrelators of Neumann (Dirichlet) 
Vasiliev  EAdS$_4$ gravity equal those of the free  (critical)  $O(N)$ model with the identification $C_2={1 / 2 N}$.  Hence GKPY duality - if correct - implies that Neumann (Dirichlet) 
Vasiliev dS$_4$ gravity is dual to the free  (critical)  $Sp(N)$ CFT$_3$.

\vspace{.5cm}

\noindent
{\bf Acknowledgements:} We are grateful to Tarek Anous, Nima Arkani-Hamed, Matthew Baumgart, Frederik Denef, Anatoly Dymarsky, Jonathan Heckman, Daniel Jafferis, Igor Klebanov, Juan Maldacena, Shiraz Minwalla, Steve Shenker, Eva Silverstein and especially  Misha Vasiliev and Xi Yin for useful discussions and correspondence. This is work is supported in part by U.S. Department of Energy grants DE-FG02-90ER40542 and DE-FG02-90ER40654. 

\appendix

\section{On-shell action formalism}\label{app:action}

In this appendix we compute the dS$_4$ Vasiliev theory boundary correlators from the point of view of the on-shell action, as discussed in \cite{Maldacena:2002vr,Harlow:2011ke}.  There is no known action formalism for the physical fields in Vasiliev theory, so we will assume that the equations of motion obey the appropriate integrability conditions to define an action, and use only the expected schematic form.

The classical quadratic action for canonically normalized higher spin fields $X$ of spins $0,2,4,\dots$, is schematically
\begin{equation}\label{bulkfree}
I_{free} \sim  \frac{1}{G_N} \int \nabla X \nabla X+ \Lambda X^2~,
\end{equation}
where as in the main text we have kept the explicit factors of $\Lambda$ but ignored other $\Lambda$-independent constants. $G_N$ here is the sole dimensionful constant needed to construct an action out of the equations of motion. The full interacting action is an expansion in $\nabla^2/\Lambda$; for example the $(2,s,s)$ interaction in (\ref{cantwoeom}) leads to the schematic form
\be\label{cantwo}
I_{2ss} \sim \frac{1}{G_N} \int \nabla^2 h X^2 + \Lambda h X^2 + \frac{1}{\Lambda} \nabla^2 h \nabla X \nabla X + h \nabla X \nabla X +\cdots\ .
\ee
Now we would like to compare the $n$-point functions in dS$_4$ to those in EAdS$_4$.  For this the schematic form of the action as an expansion in $\frac{1}{\Lambda}\nabla ^2$ is enough, we do not need the explicit interactions.

In the action formalism correlation functions with future boundary conditions are defined as derivatives of a bulk wavefunction,
\be\label{predcor}
\langle J^{(s_1)}(x_1) \cdots J^{(s_n)}(x_n) \rangle_{bulk} \sim \left.\frac{\delta^n \Psi_{bulk}[\phi_0]}{\delta \phi_0^{(s_1)}(x_1) \cdots \delta \phi_0^{(s_n)}(x_n)}\right|_{\phi_0= 0}  \ ,
\ee
where $\phi_0^{(s)}$ is the canonical data for the spin-$s$ field on a spatial slice.  Note that this is a bulk correlator with future boundary conditions of the type discussed in \cite{Anninos:2011jp}, which is not equal to a bulk expectation value in any quantum state. In the leading semiclassical approximation, the bulk wavefunction is
\be
\Psi_{bulk}[\phi_0] = e^{i I_{cl} }
\ee
where $I_{cl}$ is the classical action of the bulk theory evaluated on a solution to the equations of motion with future boundary conditions specified by $\phi_0$, up to local counterterms. Initial conditions are imposed by an $i\epsilon$ prescription that places the bulk fields in the Bunch-Davies vacuum.

For Vasiliev gravity the on-shell action $I_{cl}$ can be computed by double analytic continuation from EAdS$_4$ as  in \cite{Maldacena:2002vr,Harlow:2011ke}.  First consider the kinetic terms.  In EAdS$_4$, for a scalar of dimension $h_+$ we set $\phi(x,z) \sim \phi_0(x) z^{3-h_+} + \cdots$ near the boundary and compute the on-shell action as a functional of the sources at a cutoff surface $z=z_c \ll 1$,
\be
-I_{EAdS} \sim \frac{\ell_{AdS}^{2}}{G_N} \int d^3 k\   k^{2h_+ - 3} z_c^{2(h_+-3)} \phi_0(k)\phi_0(-k)+ \mbox{counterterms}
\ee
where `counterterms' means terms local in position space and `$\sim$' means we have dropped a real constant independent of $k,\ell_{AdS},z_c$.  Taking derivatives and Fourier transforming leads to the position space correlation function
\be
\langle \tilde{\bigO} \tilde{\bigO} \rangle \sim  \frac{\ell_{AdS}^{2}}{G_N} x^{-2h_+} z_c^{2(h_+-3)} \ .
\ee
Wavefunction renormalization $\bigO \sim z_c^{3-h_+} \tilde\bigO$ gives 
\be\label{twoads}
\langle \bigO \bigO \rangle \sim \frac{\ell_{AdS}^{2}}{G_N} x^{-2h_+} \ .
\ee
This is the tree-level answer in the EAdS$_4$/CFT$_3$ correspondence.  To transform to de Sitter space, we use the double analytic continuation (\ref{ancon}), while keeping $G_N$ fixed, and the identification $- I_{EAdS} = i I_{dS} $.  Note that the on-shell action is defined intrinsically in de Sitter; we are just using EAdS$_4$ as a tool to compute it easily. The condition of regularity in the interior of EAdS$_4$ analytically continues to the $i\varepsilon$ prescription appropriate for the Bunch-Davies vacuum.  We find
\be
i I_{dS} \sim - \frac{\ell^2_{dS}}{G_N}(-i\eta_c)^{2(h_+-3)} \int \frac{d^3 k}{(2\pi)^3} k^{2h_+-3} \phi_0(k)\phi_0(-k) + \cdots
\ee
and choosing $\bigO \sim (-i\eta_c)^{3-h_+} \tilde\bigO$ gives
\be\label{opnorm}
\langle \bigO \bigO \rangle \sim - \frac{\ell^2_{dS}}{G_N} x^{-2h_+} \ .
\ee
This differs by a sign from (\ref{twoads}), but so far this is just a choice of normalization.  The higher $n$-point functions are fixed unambiguously once we have normalized the 2-point functions in this way, and we will see that this choice is necessary to eliminate phases from $n$-point correlators in dS$_4$.  

For fields with spin, the kinetic term (\ref{bulkfree}) has $s$ additional powers of $g^{\mu\nu}$ to contract indices. Each of these comes with a factor of $\ell_{AdS}^{-2}$, which gives an additional factor of $i^{2s}$ in the analytic continuation, in addition to those obtained from the continuation of $z$.  
Denoting the EAdS$_4$ correlator by 
\be
\langle J^{(s)} J^{(s)} \rangle_{EAdS} = \frac{\ell^2_{AdS}}{G_N} f(x)
\ee
where $f(x)$ contains the position and tensor dependence, if we absorb an additional factor of $i^s$ into each de Sitter operator then we obtain
\be
\langle J^{(s)} J^{(s)} \rangle_{dS} = - \frac{\ell^2_{dS}}{G_N} f(x) \ .
\ee

Now consider interactions.  As an example, take the cubic interaction between three canonically normalized higher spin fields $X_1,X_2,X_3$ of spins $s_1,s_2,s_3$,
\be
I_{int} =  \frac{\Lambda^{1-n}}{G_N} \int \sqrt{g} \nabla ^{p} X_1 \nabla ^q X_2 \nabla ^{2n-p-q} X_3
\ee
with indices contracted in an arbitrary way and $p,q \geq 0$, $2n\geq p+q$.  To compute the tree-level correlator, we plug in solutions of the free equations of motion with fixed boundary conditions and compute $I_{int}$ as a function of $\ell_{AdS}, z_c$, and the boundary data.  The resulting function of the boundary data has the form
\be
\left( \frac{\ell^2_{AdS}}{G_N} \right) \ell_{AdS}^{-s_1 - s_2 - s_3} f(z_c; X_{10}, X_{20}, X_{30})
\ee
with $f$ some cutoff-dependent functional of the boundary data $X_{i0}$. The factors of $\ell$ come from $\Lambda = -3/\ell^2_{AdS}$ and counting powers of $g^{\mu\nu}$. Performing the analytic continuation and absorbing a factor of $i^{s}$ into each de Sitter operator as required by the normalization above, we find
\be
\langle J^{(s_1)} J^{(s_2)} J^{(s_3)} \rangle_{dS} = - \langle J^{(s_1)} J^{(s_2)} J^{(s_3)} \rangle_{EAdS} \ .
\ee
Since the action for canonical fields is always an expansion in $\frac{1}{\Lambda}\nabla ^2$, this counting will continue to hold for arbitrary interaction terms that appear in the Vasiliev theory.  Therefore this result actually holds for general tree-level correlation functions,
\be
\langle J^{(s_1)} J^{(s_2)} \cdots \rangle_{dS} = - \langle J^{(s_1)} J^{(s_2)} \cdots \rangle_{EAdS} \ .
\ee
In the bulk theory with Neumann boundary conditions on the scalar, loop corrections are required to cancel by GKPY duality.  Therefore we have reproduced the result of section \ref{ss:ancon} with Neumann boundary conditions, and these correlators agree with the $Sp(N)$ CFT at the UV fixed point.  The duality at the IR fixed point then follows by a Legendre transform of the bulk and boundary generating functionals   \cite{Klebanov:1999tb,Witten:2001ua,Gubser:2002vv,Giombi:2011ya}.

\section{Reality conditions}\label{app:realx}

In this appendix we demonstrate that the spin-$s$ component fields $X_{i_1 \cdots i_s},~~i=1,2,3$ for even $s$ are real in both EAdS$_4$ and dS$_4$. These describe all the 
physical degrees of freedom in FG gauge. The fact that the FG gauge $X$ reality condition is the same in EAdS$_4$ and dS$_4$ is required for our derivation of the central relation (\ref{rle})
between correlators.

The $X$ fields appear as  components of a master field which is subject to a reality condition which differs between EAdS$_4$ and dS$_4$. 
Reality conditions on the $X$s are implied by those on the master field.
This master field reality condition was specified for dS$_4$ in \cite{Vasiliev:1986td}. However we shall follow here \cite{Iazeolla:2007wt} (ISS) which contains the same  reality condition (in a slightly different notation)
and has many details explicitly written out. In this appendix we refer the reader directly to ISS for all conventions. 

According to ISS table 1, the dS$_4$ master field reality condition at $Z=0$ (which is sufficient to determine the $X$ reality condition) on the one form $W$  of (ISS 2.70) is, using (ISS 2.38)
\be W(x^\mu | y_\a,\bar y_{\dot \a})^\dagger=-\pi W (x^\mu| y_\a,\bar y_{\dot \a})=-W (x^\mu| -y_\a,\bar y_{\dot \a})\ee
where (ISS 2.40) $(y_\a)^\dagger= \bar y_{\dot \a}$.  The FG gauge spin $s>0$ fields 
$X_{i_1 \cdots i_s},~~i=1,2,3$ are related to the coefficient fields $W(x^\mu)_{\a_1 \cdots \a_n;\dot \a_1\cdots \a_m }$ of the $(y,\bar y)$ expansion of $W$  with $m=n=s-1$ by  \be X_{\mu i_1 \cdots i_{s-1}}dx^\mu=W_{\a_1...\a_{s-1}, \dot \a_1...\dot \a_{s-1}}
\sigma^{\a_1\dot \a_1}_{i_1}\cdots \sigma^{\a_{s-1}\dot \a_{s-1}}_{i_{s-1}} \ .\ee
Using reality of $y_\a\bar y_{\dot \a}\sigma_i^{\a\dot\a}$ and the fact that $s$ is always even, the $W$ reality condition on the physical FG gauge fields reduces to 
\be \label{dsxc} X_{i_1 \cdots i_s}={\rm real}.\ee
The EAdS$_4$ master field reality condition at $Z=0$  on  $W$  is, using  (ISS 2.38) and (ISS 2.42) 
\be W(x^\mu | y_\a,\bar y_{\dot \a})^\dagger=-\rho \pi W (x^\mu| y_\a,\bar y_{\dot \a})=-W (x^\mu| -y_\a^\dagger,\bar y_{\dot \a}^\dagger)\ee
where here $\dagger$ is complex conjugation.  For even $s$ we again  recover (\ref{dsxc}) as desired.

\begin{singlespace}

\end{singlespace}

\end{document}